# PIE: A Domain-Specific Language for Interactive Software Development Pipelines

Gabriël Konat[a], Michael J. Steindorfer[a], Sebastian Erdweg[a], and Eelco Visser[a]

a   Delft University of Technology, The Netherlands

**Abstract**

**Context.** Software development pipelines are used for automating essential parts of software engineering processes, such as build automation and continuous integration testing. In particular, *interactive* pipelines, which process events in a live environment such as an IDE, require *timely* results for low-latency feedback, and *persistence* to retain low-latency feedback between restarts.

**Inquiry.** Developing an incrementalized and persistent version of a pipeline is one way to reduce feedback latency, but requires implementation of dependency tracking, cache invalidation, and other complicated and error-prone techniques. Therefore, interactivity complicates pipeline development if timeliness and persistence become responsibilities of the pipeline programmer, rather than being supported by the underlying system. Systems for programming incremental and persistent pipelines exist, but do not focus on ease of development, requiring a high degree of boilerplate, increasing development and maintenance effort.

**Approach.** We develop Pipelines for Interactive Environments (PIE), a Domain-Specific Language (DSL), API, and runtime for developing interactive software development pipelines, where ease of development *is* a focus. The PIE DSL is a statically typed and lexically scoped language. PIE programs are compiled to programs implementing the API, which the PIE runtime executes in an incremental and persistent way.

**Knowledge.** PIE provides a straightforward programming model that enables direct and concise expression of pipelines without boilerplate, reducing the development and maintenance effort of pipelines. Compiled pipeline programs can be embedded into interactive environments such as code editors and IDEs, enabling timely feedback at a low cost.

**Grounding.** Compared to the state of the art, PIE reduces the code required to express an interactive pipeline by a factor of 6 in a case study on syntax-aware editors. Furthermore, we evaluate PIE in two case studies of complex interactive software development scenarios, demonstrating that PIE can handle complex interactive pipelines in a straightforward and concise way.

**Importance.** Interactive pipelines are complicated software artifacts that power many important systems such as continuous feedback cycles in IDEs and code editors, and live language development in language workbenches. New pipelines, and evolution of existing pipelines, is frequently necessary. Therefore, a system for easily developing and maintaining interactive pipelines, such as PIE, is important.

## The Art, Science, and Engineering of Programming

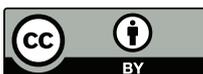

## 1 Introduction

A pipeline is a directed acyclic graph of processors in which data flows from the output of one processor to the input of its succeeding processors. Pipelines are ubiquitously used in computer hardware and software. E.g., in hardware, CPUs contain instruction pipelines that allow interleaved execution of multiple instructions that are split into fixed stages. Software pipelines compose software components by programmatically connecting their input and output ports (e.g., UNIX pipes).

In software development, pipelines are used to automate parts of the software engineering process, such as building software systems via build scripts, or continuously testing and integrating the composition of subsystems. Such pipelines are suitable for batch-processing, and often run isolated on remote servers without user interaction.

*Interactive* software development pipelines build software artifacts, but react instantly to changes in input data and provide timely feedback to the user. Typical examples are continuous editing of source code in an Integrated Development Environment (IDE), providing feedback through editor services such as syntax highlighting; selective re-execution of failing test cases in the interactive mode of a build system during development; or development of languages in a language workbench [14].

Interactive pipelines focus on delivering *timely* results when processing an event, such that the user can subsequently act on the results. Furthermore, an interactive software development pipeline should *persist* its state on non-volatile memory so that a session can be restarted without re-execution. Especially in the context of an IDE, restarting the development environment should not trigger re-execution of the entire pipeline, especially if pipeline steps are costly, such as advanced static analyses [42].

Interactivity complicates the development of pipelines, if *timeliness* and *persistency* become responsibilities of the pipeline programmer, rather than being supported by the underlying system. Developing an incrementalized version of an expensive operation is one way to reduce the turnaround time when re-executing the operation. However, implementing support for incrementality in a pipeline is typically complicated and error-prone. Similarly, persisting the result of expensive operations reduces the turnaround time when restarting a session, but requires tedious management of files or a database. Furthermore, when persistency is combined with incrementality, dependency tracking and invalidation is required, which is also complicated and error-prone. Therefore, an expressive system for easily developing correct incremental and persistent interactive software development pipelines is required.

One system that partially achieves this is Pluto [12], a sound and optimal incremental build system. Pluto supports dynamic dependencies, meaning that dependencies to files and other build steps are created during build execution (as opposed to before or after building), enabling both increased incrementality through finer-grained dependencies, and increased expressiveness. While Pluto focusses on build systems, it is well suited for expressing correct incremental and persistent pipelines. However, ease of development is not a focus of Pluto, as pipelines are implemented as Java classes, requiring significant boilerplate which leads to an increase in development and maintenance effort. Furthermore, persistence in Pluto is not fully automated because pipeline developers need to manually thread objects through pipelines to prevent hidden dependencies, and domain-specific features such as file operations are not first class. These are open problems that we would like to address.

In this paper, we introduce Pipelines for Interactive Environments (PIE), a Domain-Specific Language (DSL), Application Program Interface (API), and runtime for programming interactive software development pipelines, where ease of development *is* a focus. The PIE DSL provides a straightforward programming model that enables direct and concise expression of pipelines, without the boilerplate of encoding incrementality and persistence in a general-purpose language, reducing development and maintenance effort. The PIE compiler transforms high-level pipeline programs into programs implementing the PIE API, resulting in pipeline programs that can be incrementally executed and persisted to non-volatile memory to survive restarts with the PIE runtime. Compiled pipeline programs can be embedded in an interactive environment such as an IDE, combining coarse grained build operations with fine-grained event processing.

To summarize, the paper makes the following contributions:

- The PIE language, a DSL with high-level abstractions for developing interactive software development pipelines without boilerplate.
- The PIE API for implementing foreign pipeline functions, and as a compilation target for the DSL, with reduced boilerplate.
- The PIE runtime that executes pipelines implemented in the API in an incremental and persistent way, which fully automates persistence and automatically infers hidden dependencies.
- An evaluation of PIE in two critical case studies: (1) modeling of the pipeline of a language workbench in an IDE setting, and (2) a pipeline for incremental performance testing.

The PIE implementation is available as open source software [26].

**Outline**  The paper continues as follows. In Section 2 we describe requirements for interactive software development pipelines, review the state of the art, and list open problems. In Section 3 we illustrate PIE by example. In Section 4 we describe the PIE API and runtime. In Section 5 we describe the syntax, static semantics, and compilation of the PIE DSL in more detail. In Sections 6 and 7 we present critical case studies of the application of PIE in an interactive language workbench and an interactive benchmarking setting. In Section 8 we discuss related work. In Section 9 we discuss directions for future work. Finally, we conclude in Section 10.

## 2 Problem Analysis

In this section, we first describe requirements for interactive software development pipelines, review the state of the art, and list open problems.

**Requirements**  We first describe the requirements for interactive software development pipelines. In order to do so, we use the example pipeline from Figure 1 as the

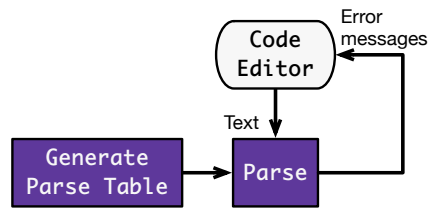

■ **Figure 1**  Example of an interactive software development pipeline, where text from a code editor is parsed, and parse error messages are displayed in the editor.

running example in this section. In this pipeline, a code editor parses its text buffers in order to display error messages interactively to the programmer. Parsing requires a parse table, which is generated by an external process and may change when a new version of a language is deployed, with new syntax that requires regeneration of the parse table. We identify the following requirements for interactive software development pipelines:

- Incrementality. A pipeline should attempt to recompute only what has been affected by a change. For example, when only a text buffer in the code editor changes, the pipeline reparses the text and new error messages are displayed, but the generated parse table is reused because it did not change.

- Correctness. Incremental pipeline executions must have the same results as from-scratch batch executions. For example, if the parse table *does* change, the pipeline also reparses text and displays new error messages.

- Persistence. Results of computation should be persisted to disk in order to enable incrementality after a restart of the pipeline. For example, if we restart the code editor, the parse table is retrieved from disk instead of requiring a lengthy recomputation.

- Expressiveness. In practice, pipelines are a lot more complex than the simple example shown here. It should be possible to express more complex pipelines as well.

- Ease of development. Pipelines are complex pieces of software, especially when the previous requirements are involved. Therefore, the development and maintenance effort of pipelines should be low.

**State of the Art**   We now review the state of the art in interactive software development pipelines, and determine to what extent existing tools meet the requirements, focussing on build systems.

Make [38], and systems with similar dependency management (e.g., Ninja, SCons, MSBuild, CloudMake, Ant), are tools for developing build systems based on declarative rules operating on files. These tools support incremental builds, but incrementality is limited to static file dependencies which are specified up front in the build rules. Because dependencies cannot be the result of computation, the dependencies must either be soundly overapproximated, which limits incrementality, or underapproximated, which is unsound. For example, a Makefile that determines the version of a Java source file, in order to parse it with the corresponding parse table file, must

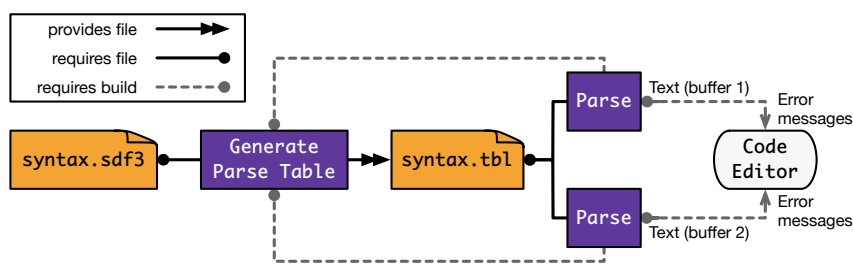

**◼ Figure 2** Pluto dependency graph created by executing the pipeline from Figure 1, where the code editor has 2 open text buffers.

depend on all parse table files instead of a single one. Therefore, a system that supports a more expressive dependency mechanism is required. A detailed discussion of dependency expressiveness can be found in Section 8, but in this section, we focus on the system with the highest dependency expressivity: Pluto.

Pluto [11] is a sound and optimal incremental build system with support for dynamic dependencies. A build system in Pluto is implemented in terms of *builders*, which are functions that perform arbitrary computations and dynamically record dependencies to files and other builders during execution. Executing a builder with an input produces a *build*, containing an output object and recorded dependencies.

Figure 2 illustrates the dependency graph Pluto produces when it executes the pipeline of Figure 1 where the code editor has two open text buffers. The dependency graph differs from the pipeline by containing builds (function calls) instead of builders (function definitions). For example, the pipeline has one parse builder, but two parse builds, one for each text buffer. We use this dependency graph to illustrate Pluto's adherence to requirements for interactive pipelines.

- Incrementality. The code editor has two text buffers open, which have separate dependencies to a parse build. When one text buffer changes, only the corresponding parse build is recomputed. Therefore, Pluto supports fine-grained incrementality.

- Correctness. The parse builds depend on the parse table build, such that when the parse table is regenerated, the parse builds are re-executed, and new error messages are displayed in the editor. Pluto enforces this by performing *hidden dependency detection*. That is, if a build requires a file, without requiring the build that provides that file, Pluto marks this as an error and aborts execution.

- Persistence. While not shown in the dependency graph, builds are persisted to disk to survive restarts.

- Expressiveness. Dependencies are recorded during build execution, allowing builds to depend on files or call other builds, based on results of computation. For example, when parsing Java code, the parse builder may choose to depend on a different parse table, based on whether we want to parse text of version 8 or 9 of Java. This greatly increases the expressiveness required for interactive software development pipelines.

- Ease of development. Pluto build systems are implemented in Java, requiring significant boilerplate.

■ **Listing 1** The parsing pipeline implemented as Java classes in Pluto.

```java
class GenerateTable extends Builder<File, Out<File>> {
  static BuilderFactory<File, Out<File>, GenerateTable> factory =
    BuilderFactoryFactory.of(GenerateTable.class, File.class);
  GenerateTable(File syntaxFile) { super(syntaxFile); }
  @Override File persistentPath(File syntaxFile) {
    return new File("generate-table-" + hash(syntaxFile));
  }
  @Override Out<File> build(File syntaxFile) throws IOException {
    require(syntaxFile); File tblFile = generateTable(syntaxFile);
    provide(tblFile); return OutputPersisted.of(tblFile);
} }
class Parse extends Builder<Parse.Input, Out<ParseResult>> {
  static class Input implements Serializable {
    File tblFile; String text; BuildRequest tblReq;
    Input(File tblFile, String text, BuildRequest tblReq) {
      this.tblFile = tblFile; this.text = text; this.tblReq = tblReq;
    }
    boolean equals(Object o) {/* omitted */} int hashCode() {/* omitted */}
  }
  @Override Out<ParseResult> build(Input input) throws IOException {
    requireBuild(input.tblReq); require(input.tblFile);
    return OutputPersisted.of(parse(input.tblFile, input.text));
} } /* ... other required code omitted ... */
class UpdateEditor extends Builder<String, Out<ParseResult>> {
  @Override Out<ParseResult> build(String text) throws IOException {
    File syntaxFile = new File("syntax.sdf3");
    File tblFile = requireBuild(GenerateTable.factory, syntaxFile).val;
    BuildRequest tblReq = new BuildRequest(GenerateTable.factory, syntaxFile);
    return requireBuild(Parse.factory, new Parse.Input(tblReq, tblFile, text));
} } /* ... other required code omitted ... */
```

To summarize, Pluto provides a great foundation for implementing interactive software development pipelines, but does not cater to the pipeline developer because ease of development is not a focus, leading to a higher implementation and maintenance effort than necessary.

**Open Problems**  The main problem is that Pluto build systems are not easy to develop. We list four concrete open problems:

- Boilerplate. Pipelines in Pluto are written in Java, which has a rigid and verbose syntax, requiring significant *boilerplate*. Pipelines are implemented as classes extending the `Builder` abstract class, as seen in Listing 1. Such a class requires generics for specifying the input and output type, a `factory` and constructor enabling other builders to create instances of this builder to execute it, a `persistentPath` method for persistence, and finally a `build` method that performs the actual build computation. The `Parse` builder requires an inner class for representing multiple input values, which must correctly implement `equals` and `hashCode`, which Pluto uses to detect if an input has changed for incrementality. Finally, calling other builders through `requireBuild` is verbose, because the `factory` is referenced, and the result is unwrapped with `.val`.

- Semi-automated persistence. Pipeline developers need to implement the `persistentPath` method of a builder and return a *unique* and *deterministic* filesystem path where the result of the builder and its input are persisted. It must be unique to prevent overlap with other builders or other inputs. For example, if the `Parse` builder persists results to the same file for different text buffers, it overwrites the persisted result of other builds. It must be deterministic such that the persisted file can later be found again. Since the OS filesystem is used for persistence, there are also limitations to which characters can be used in paths, and to how long a path can be. For example, on Windows, the current practical limit is 260 characters which is frequently reached with deeply nested paths, causing persistence to fail.

- Hidden dependencies. Hidden dependency detection is crucial for sound incremental builds, but is also cumbersome. In the pipeline in Listing 1, we must construct a build request object for the parse table generator, pass that object to the `Parse` builder, and require it to depend on the parse table generator. This becomes tedious especially in larger and more complicated pipelines.

- Missing domain-specific features. Path (handle to file or directory) and list operations, which are prevalent in software development pipelines, are not first class in general-purpose languages such as Java.

Solving these concrete problems requires a proper abstraction over interactive software development pipelines, which we present in subsequent sections.

## 3  PIE by Example

To solve the open problems from the previous section, we introduce PIE: a DSL, API, and runtime for developing and executing interactive software development pipelines. Pipelines in PIE have minimal boilerplate, fully automated persistence, automatically infer hidden dependencies, and have domain-specific features such as path and list operations. In this section we illustrate PIE by means of an example that combines building and interaction. We discuss the example and the requirements for this pipeline, present the pipeline in the PIE DSL, and discuss its features and execution.

**Example Pipeline: Syntax-Aware Editors**  As example we consider a code editor with syntax styling based on a syntax definition. The pipeline to support this use case is depicted by the diagram in Figure 3. It generates a parse table from a syntax definition, parses the program text of an editor, computes syntax styling for each token, and finally applies the computed syntax styling to the text in the editor. We want this pipeline to be *interactive* by embedding it into the IDE such that changes to the syntax definition as well as changes to the text in an editor are reflected in updates to syntax styling. The example in Figure 3 is representative for *language workbenches* [13, 14],

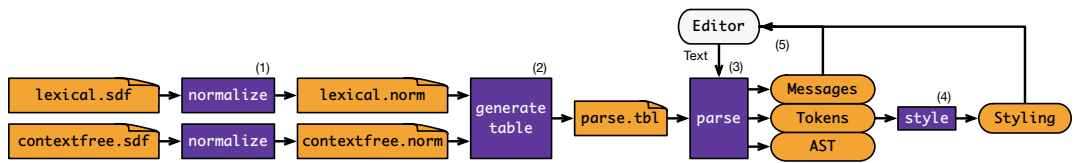

■ **Figure 3**   The example pipeline: (1) normalization of SDF syntax definition source modules, (2) generation of a parse table from the normalized modules comprising the definition for a language, (3) parsing the text of an editor using the parse table, (4) computing the styles for the parsed tokens, and (5) displaying the styling and error messages in an editor.

which support edits on a language definition that are immediately reflected in the programming environment that is derived from it.

Concretely, we instantiate the pipeline with components from the Spoofax language workbench [24]. We process an SDF [45, 46] syntax definition in two stages. First, syntax definition modules are separately transformed (normalized) to a core language. Next, the normalized modules comprising the syntax definition for a language are transformed to a parse table. The parse table is interpreted by a scannerless parser [6, 44] to parse the contents of an editor, returning an Abstract Syntax Tree (AST), token stream, and error messages. A syntax highlighter annotates tokens in the token stream with styles.

**Integrated Pipelines with the PIE DSL**   Listing 2 shows the pipeline program in the PIE DSL. We first explain what each function does, and then discuss the features and execution of PIE in more detail.

The `normalize` function executes a command-line tool to normalize an SDF source file into a normalized version that is ready for parse table generation, and retrieves (dynamic) dependencies from the generated dependency (.dep) file, implemented by the `extract-deps` foreign function. The `generate-table` function executes a command-line tool on normalized files, creating a parse table file. The `parse` function, when given a parse table object, parses text into an AST, token stream, and error messages. The `style` function produces a styling based on a token stream, which can be used in source code editors for styling the text of the source code. Finally, the `update-editor` function defines the complete pipeline by composing all previously defined functions.

**Composing Pipelines with Functions**   In PIE, pipelines are defined in terms of function definitions which are the reusable processors of the pipeline, and function calls that compose these processors to form a pipeline. Function calls register a dynamic *call dependency* from caller to callee.

**Domain-Specific Types and Dependencies**   Since build pipelines often interact with files and directories, PIE has native support for `path` types and several operations on paths. Path literals such as `./lexical.sdf` provide an easy way to instantiate relative or absolute paths. The `requires` operation dynamically registers a *path dependency* from the current function call to the path, indicating that the function reads the path,

■ **Listing 2**  PIE DSL program for the pipeline illustrated in Figure 3. Identifiers of foreign functions and data types are omitted for brevity.

```
func normalize(file: path, includeDirs: path*) -> path = {
  requires file; [requires dir with extension "sdf" | dir <- includeDirs];
  val normFile = file.replaceExtension("norm");
  val depFile = file.replaceExtension("dep");
  exec(["sdf2normalized"] + "$file" + ["-I$dir" | dir <- includeDirs] +
    "-o$normFile" + "-d$depFile");
  [requires dep by hash | dep <- extract-deps(depFile)];
  generates normFile; normFile
}
func extract-deps(depFile: path) -> path* = foreign
func generate-table(normFiles: path*, outputFile: path) -> path = {
  [requires file by hash | file <- normFiles];
  exec(["sdf2table"] + ["$file" | file <- normFiles] + "-o$outputFile");
  generates outputFile; outputFile
}
func exec(arguments: string*) -> (string, string) = foreign

data Ast = foreign {} data Token = foreign {} data Msg = foreign {}
data ParseTable = foreign {} data Styling = foreign {}
func table2object(text: string) -> ParseTable = foreign
func parse(text: string, table: ParseTable) -> (Ast, Token*, Msg*) = foreign
func style(tokenStream: Token*) -> Styling = foreign
func update-editor(text: string) -> (Styling, Msg*) = {
  val sdfFiles = [./lexical.sdf, ./contextfree.sdf];
  val normFiles = [normalize(file, [./include]) | file <- sdfFiles];
  val parseTableFile = generate-table(normFiles, ./parse.tbl);
  val (ast, tokenStream, msgs) = parse(text, table2object(read parseTableFile));
  (style(tokenStream), msgs)
}
```

whereas `generates` records a dependency indicating the function creates or writes to the path. The `read` operation reads the text of a given path, and also registers a path dependency.

Path dependencies to directories can specify a filter such as `with extension "sdf"` to only create dependencies to files inside the directory that match the filter. Finally, path dependencies can specify how changes are detected. For example `requires dep by hash` indicates that a change is only detected when the hash of the file changes, instead of the (default) modification date, providing more fine grained dependency tracking.

**Foreign Functions and Types**   Some functions are `foreign`, indicating that they are implemented outside of the PIE DSL, either because they are outside of the scope of the DSL (e.g., text processing required for `extract-deps`), or because they require system calls. For example, `exec` is a foreign function that takes a list of command-line arguments, executes a process with those arguments, and returns its standard output and error text. Unlike `read`, `exec` is not first class, because it does not induce special (path) dependencies.

PIE contains several built-in types such as `string` and `bool`, but foreign types can be defined to interface with existing types. Foreign data types are required to integrate with existing code, such as an editor that expects objects of type `Styling` and `Msg`, returned by foreign functions `parse` and `style`.

**Comprehensions**    Pipelines frequently work with lists, which are natively supported in PIE by annotating a type with `*` multiplicity. Lists are instantiated with list literals between `[ ]`, and concatenated using `+`. List comprehensions such as `[f(elem) | elem <- elems]` transform a list into a new list by applying a function `f` to each element of the list.

**Execution and IDE Integration**    To execute a pipeline, we compile it into a program implementing the PIE API. We embed the compiled pipeline, together with the PIE runtime, into an IDE such as Eclipse. When an editor in Eclipse is opened or changed, it calls the `update-editor` function through the PIE runtime with the text from the editor. The PIE runtime then incrementally executes (and persists the results of) the pipeline and returns `Styling` and `Msg`s objects, which Eclipse displays in the editor. Because the results of the pipeline are persisted, a restart of Eclipse does not require re-execution of the pipeline. This becomes especially important with larger pipelines.

**Solutions to Open Problems**    PIE solves the open problems listed in Section 2. First of all, PIE minimizes boilerplate by enabling direct expression of pipelines in the PIE DSL through function definitions, function calls, foreign data types and functions, and path dependencies. The compiler of the DSL generates the corresponding boilerplate. Furthermore, PIE supports fully automated persistence. There is no need to specify where the result of a function call is stored. The PIE runtime stores results automatically based on the function name and input arguments. It persists the function arguments, return value, and dependencies in a key-value store, preventing filesystem issues.

Hidden call dependencies are automatically inferred. In other words, when a function requires files that are generated by another function, the first function does not explicitly need to call the latter. For example, `generate-table requires` files that `normalize generates`, but does not need to explicitly call `normalize` to record a call dependency which keeps the required files up-to-date. The PIE runtime infers these dependencies by keeping track of which function call generated a file, further reducing boilerplate. Note that this only infers *call dependencies*, not path dependencies, which still need to be declared by the pipeline programmer.

Finally, the PIE DSL caters to the pipeline developer by including domain-specific features – such as path type and operations, list type and comprehensions, string and path interpolation, and tuples – to make pipeline development convenient.

Solving these problems reduces the implementation and maintenance effort. The equivalent Pluto implementation for this pipeline requires 396 lines of Java code in 8 files (excluding comments and newlines), whereas the PIE implementation is

■ **Listing 3**  The PIE `Func` API and implementation of a parsing pipeline in that API.

```
typealias In = Serializable; typealias Out = Serializable
interface Func<in I:In, out O:Out> {
  fun ExecContext.exec(input: I): O
}
interface ExecContext {
  fun <I:In, O:Out, F:Func<I, O>> requireCall(clazz: KClass<F>, input: I,
    stamper: OutputStamper = OutputStampers.equals): O
  fun require(path: PPath, stamper: PathStamper = PathStampers.modified)
  fun generate(path: PPath, stamper: PathStamper = PathStampers.hash)
}

class GenerateTable: Func<PPath, PPath> {
  override fun ExecContext.exec(syntaxFile: PPath): PPath {
    require(syntaxFile); val tableFile = generateTable(syntaxFile);
    generate(tableFile); return tableFile
} }
class Parse: Func<Parse.Input, ParseResult> {
  data class Input(val tableFile: PPath, val text: String): Serializable
  override fun ExecContext.exec(input: Input): ParseResult {
    require(input.tableFile); return parse(input.tableFile, input.text)
} }
class UpdateEditor: Func<String, ParseResult> {
  override fun ExecContext.exec(text: String): ParseResult {
    val tableFile = requireCall(GenerateTable::class, path("syntax.sdf3"))
    return requireCall(Parse::class, Parse.Input(tableFile, text))
} }
```

over 6 times shorter by only requiring 62 lines of code in 2 files. The PIE code consists of 34 lines of PIE DSL code, and 28 lines of PIE API code for interfacing with foreign functions.

## 4 PIE API and Runtime

In this section, we review the PIE API and runtime, and our reasons for not directly reusing the Pluto runtime.

**API** The PIE API is a Kotlin [23] library for implementing PIE function definitions on the Java Virtual Machine (JVM). Kotlin is a programming language with a focus on reducing verbosity and increasing extensibility compared to Java, while maintaining fully compatible with Java by running on the JVM. It shares many goals with Scala [10], but additionally focusses on fast compile times and simplicity. We chose to specify the API in Kotlin instead of Java, because it has a more flexible and concise syntax. The PIE API is heavily based on the Pluto API, but uses terminology from the pipeline domain (functions instead of builders), and requires less boilerplate.

Listing 3 illustrates the parsing pipeline implemented in the PIE API. A pipeline function definition is implemented by creating a class which subtypes the `Func` inter-face and overrides the `exec` function. The `exec` function takes an input, is executed in an execution context `ExecContext`, and produces an output. The execution context enables calling other pipeline functions through the `requireCall` function, and recording of path dependencies through the `require` and `generate` functions, using Kotlin's extension functions to make these functions accessible without a qualifier. The PIE runtime uses this execution context for dependency tracking and hidden dependency inference.

Inputs of `Func` implementations must be immutable, `Serializable`, and have an `equals` and `hashCode` implementation. These properties are required so that the PIE runtime can assume objects do not change inside a cache, can persist objects to non-volatile memory, and can detect if an object has changed for incrementality. The types used in Listing 3 all adhere to these properties. Furthermore, Kotlin's data classes automatically implement `equals` and `hashCode`, reducing boilerplate for multiple input arguments.

Outputs of functions must adhere to the same properties, with the exception that outputs can opt-out of serialization. Some outputs are in-memory object representations and cannot be serialized, are too large to be serialized, or are not immutable. PIE supports these kind of objects as outputs of function calls, by wrapping the output in a special class (`OutTransient`) which prevents serialization. PIE still caches these outputs in volatile memory. However, when the runtime is restarted (thus clearing the in-memory cache), and such an output is requested by calling the function, PIE re-executes the function to recreate the output.

Although it is possible to implement a full pipeline directly in this API, there is more boilerplate involved compared to writing the pipeline in the PIE DSL. Therefore, the API should only be used for implementing foreign functions, such as interfacing with a parse table generator and parser, or for executing system calls such as executing command-line tools. However, reduced boilerplate for implementing foreign functions reduces implementation and maintenance effort.

**Runtime**   The job of the PIE runtime is to execute a pipeline – represented as a set of `Func` implementations, from compiled PIE DSL code, and from foreign function implementations against the PIE API – in an incremental and persistent way. The runtime is largely based on the Pluto runtime, from which we inherit the sound and optimal incremental and persistent build algorithm. However, we incorporate fully automated persistence and hidden dependency inference in the PIE runtime.

The runtime calls a `Func` by calling its `exec` function with an input argument, under an execution context. During execution, a function may call other functions, and record path dependencies, through the execution context, and finally return a value. After a function has been executed, the runtime persists the returned value and recorded dependency information in a key-value store, by mapping the function call (`Func` instance and input argument) to the returned value and dependency information. This mapping is used by the incremental build algorithm as a cache and for retrieving dependency information. We use the LMDB [41] key-value database, which persists to a single file on the filesystem, and is memory-mapped for fast read access.

Therefore, we fully automate persistence, meaning that pipeline developers are freed from reasoning about persistence.

To infer hidden dependencies, whenever a path (handle to file or directory) is generated, the runtime maps (in the key-value store) the path to the function call that generated the path. Whenever a path is required, the runtime consults the mapping to look up if that path was generated by a function call. If it was, then a function call dependency is inferred from the current executing function call to the function call that generated the path. For example, in Listing 3, a call of `GenerateTable` generates the parse table file, which a call of `Parse` requires. The runtime then infers a dependency from the `Parse` call to the `GenerateTable` call. This is sound, because there may be at most one function call that generates a single path. We validate this property and abort execution when multiple function calls generate a single path. Therefore, we automatically infer hidden dependencies.

**Reusing the Pluto Runtime**    We have implemented our own API and runtime, instead of reusing the Pluto runtime, for the following three reasons. First of all, we reimplemented parts of the Pluto runtime in order to better understand Pluto's incremental rebuild algorithm and concrete implementation. Second, we wanted to reduce boilerplate for writing foreign functions. Third, automated persistence would be hard to implement in Pluto, because Pluto requires every pipeline function to implement a `persistentPath` function (as seen in Listing 1), which returns a unique filesystem path for persisting the result of executing a function with a particular input. We could generate a `persistentPath` implementation from the PIE DSL, but then foreign functions still need to manually implement this functions. Furthermore, filesystem paths may not contain certain characters, and have size limits (e.g., 260 characters on many Windows systems), which makes using files as a persistent storage complicated and error prone. Therefore, in the PIE runtime, we persist to a memory-mapped database.

## 5    PIE Language

In this section, we present PIE's language definition. We present PIE's syntax specification, describe domain-specific language constructs, and briefly look at static semantics. Finally, we describe compilation from the PIE language to the API, providing incremental and persistent pipeline execution when executed with the PIE runtime.

**Syntax**    Listing 4 shows PIE's syntax through an EBNF grammar specification. PIE programs are composed of (foreign) function definitions and foreign data types at the top level. Its constructs can be categorized into base constructs that can be directly translated to a general purpose language, and special constructs for the domain of interactive software development pipelines that require a special translation. Base constructs include regular unary and binary operations, control flow, list comprehensions, value declarations and references, function definitions and calls, early return or failure, literals, and string interpolation. Special constructs include path types, path

■ **Listing 4** PIE's syntax definition in a dialect of EBNF.

```
idchr = ?[a-zA-Z0-9-_]?; id = {idchr}; qid = {idchr|"."}; int = ["-"]{?[0-9]?};
(* top-level definitions *)
func_head = id "(" {id ":" t, ","} ")" "->" t;
func_def  = "func" func_head "=" ("foreign" id | "foreign java" qid "#" id | e);
data_def  = "data" id [":" id] "foreign java" id "{" {"func" func_head "}";
program   = {func_def | data_def};
(* types and expressions *)
t = "unit"|"bool"|"int"|"string"|"path" | id | t "?" | t "*" | "(" {t, ","} ")";
e = "{" {e, ";"} "}" | "(" e ")"
  | "!" e | e "!" | e ("==" | "!=" | "||" | "&&" | "+") e
  | "if" "(" e ")" e ["else" e] | "[" e "|" binder "<-" e "]" | "val" binder "=" e
  | id | id "(" {e, ","} ")" | e "." id "(" {e, ","} ")"
  | "requires" e ["with" filter] ["by" stamper] | "generates" e ["by" stamper]
  | "exists" e | "read" e | "list" e ["with" filter] | "walk" e ["with" filter]
  | "return" e | "fail" e
  | "unit" | "true" | "false" | int | "null" | "(" {e, ","} ")" | "[" {e, ","} "]"
  | '"' {?~[\"\$\n\r]? | '\\$' | '\\"' | "$" id | "${" e "}"} '"'
  | ["."] "/" {?~[\n\r\$\,\;\]\)\ ]? | '\\' | '\\$' | "$" id | "${" e "}"};
binder = bind | "(" {bind, ","} ")"; bind = id | id ":" t;
filter = ("regex" | "pattern" ["s"] | "extension" ["s"]) e;
stamper = "exists" | "modified" | "hash";
```

literals, dependencies (`requires` and `generates`), and operations (`exists`, `read`, `list`, and `walk`); foreign function definitions and calls; and foreign data definitions.

We intentionally keep PIE's constructs simple in order to support incrementality and persistence, with concise expression of pipelines, while still supporting a wide range of different pipelines. For example, PIE does not allow assignment or other forms of mutation, because mutation complicates incrementality support. Instead, immutability allows the dynamic semantics to perform caching for improved incrementality.

**Static Semantics**   PIE is a statically typed and lexically scoped language. As base types, PIE has the unit type, booleans, integer, strings, paths, and user-defined foreign data types. Types can be made optional (`t?`), into a list (`t*`), and composed into tuples (`(t1, t2)`). All data type and function definitions are explicitly typed, but types are inferred inside function bodies. Static type checks prevent mistakes in the pipeline from appearing at runtime. For example, it is not possible to call a pipeline function with an argument of the wrong type, as PIE's type checker will correctly mark this as a type error. Name binding prevents mistakes such as duplicate definitions and unresolved references.

**Compilation**   To execute a PIE program with the PIE runtime, we compile it to a Kotlin program implementing the PIE API. We compile every function definition in the program to a class implementing `Func`, with corresponding input and output types, and compile its function to the `exec` method. Multiple function arguments, as well as tuple types, are translated into an immutable data class, implementing the required `equals`, and `hashCode` functions, and the `Serializable` interface. Function calls are compiled to `requireOutput` calls on the execution context, which records a function call dependency and incrementally executes that function.

Path dependencies are translated to `require` and `generate` calls on the execution context, which records path dependencies, and which infers hidden dependencies when requiring a generated file. Path dependencies can use different *stampers*, which instruct the PIE runtime as to how generated and required paths are checked for changes during incremental execution. The `exists` stamper checks that a file or directory exists, `modified` compares the modification date of a file or directory, and `hash` compares the hash of a file, or the hashes of all files in a directory. The `exists`, `read`, `list`, and `walk` path operations are translated to function calls of built-in functions that perform these tasks and register the corresponding path dependencies. For example, the `walk` construct recursively walks over files and directories in a top-down fashion, returns them, and registers dependencies for each visited directory. Some path constructs also accept a *filter* which filter down which files and directories are visited. For example, a `requires` on a directory with a filter only creates path dependencies for files and directories which are accepted by the filter. A regular expression, ANT pattern, or file extension filter can be used.

Other constructs (ones that do not affect incrementality or persistence) are compiled directly to Kotlin expressions. For example, list comprehensions are translated to `maps`.

## 6    Case Study: Spoofax Language Workbench

We evaluate PIE using two critical [15] case studies that are representative for the domain of interactive software development pipelines. In this section we discuss a case study in the domain of language workbenches. In the next section we discuss a case study in the domain of benchmarking.

Spoofax [24] is a language workbench for developing textual programming languages. Spoofax supports simultaneous development of a language definition and testing the programming environment generated from that language definition. This requires complex pipelines, including bootstrapping of languages [28]. In this case study we evaluate the feasibility of implementing the Spoofax pipeline using PIE.

In the Spoofax ecosystem, a programming language is specified in terms of multiple high-level declarative meta-language definitions, where each meta-language covers a language-independent aspect (e.g., separate syntax definition [45], name binding rules [3, 29, 35], or the dynamic semantics definition of a programming language [43]). Subsequently, Spoofax generates a complete implementation of a programming language, given all the meta-language definitions. Dividing a programming language implementation into linguistic abstractions in terms high-level meta-language definitions is the key enabler for maintainability of a language, however it complicates the necessary (interactive) software development pipelines.

Spoofax supports interactive language development in the Eclipse IDE, including developing multiple language specifications side-by-side. In contrast to a regular IDE that solely processes changes of source files in the source language, Spoofax additionally comes with support for interactive software development pipelines that respond to language specification changes. For example, changes to the syntax specification are reflected by reparsing source files of the language. In order to achieve this goal, Spoofax will: (1) execute a pipeline to regenerate the language implementation based on the language specification, (2) reload the updated language implementation into the language registry, and (3) execute a pipeline for all open source files of the changed language.

The pipeline for source files will: (1) parse the source file into an AST and token stream, (2) generate syntax styling based on the token stream, (3) show parse errors (if any) and apply syntax styling, and (4) analyze and transform the source file.

### 6.1 Pipeline Re-Implementation

We have implemented Spoofax' management of multiple languages, parsing, and syntax-based styling with the PIE pipeline that is illustrated in Listing 5. This is an extension to the example pipeline of Section 3, but is still a subset of the complete pipeline due to space constraints. We omit the `foreign` keyword for brevity.

**Language Specification Management**   The first part of the pipeline is used to manage multiple language specifications. The `LangSpec` data type represents a language specification, which has a file extension and configuration required for syntax specification and styling. The `Workspace` type represents a workspace with multiple language specifications, which has a list of relevant file extensions, and a function to get the `LangSpec` for a `path` based on its extension. The aforementioned data types are similar to classes by binding function definitions to them. In this particular case their implementations are foreign (i.e., implemented in a JVM language), but registered in PIE in order for using them in an interactive software development pipeline. An instance of the `Workspace` (which contains `LangSpec`s) is created by the `getWorkspace` function from a configuration file. Interfacing with foreign functions and data types is a key enabler for embedding PIE pipelines in other programs, while still benefiting from domain-specific features such as dependency tracking.

**Parse Table Generation, Parsing, and Styling**   The second part implements parsing. There are several foreign data and function definitions which bind to Spoofax's tools. For example, `sdf2table` takes a specification in the SDF meta-language, and produces a `ParseTable` which can be used to parse programs with the `jsglrParse` function. The `parse` function takes as input the text to parse and the language specification containing the syntax specification `mainFile` to derive a parser from, creates a parse table for the language specification, and uses that to parse the input `text`. Parsing returns a product type containing the `Ast`, `Tokens`, and error `Messages`. Since parsing can fail, the AST and tokens are annotated with ? multiplicity to indicate that they are nullable (optional). The third part implements syntax-based styling, similarly to parsing.

■ **Listing 5** Spoofax pipeline in PIE, with support for developing multiple language specifications, parsing, syntax styling, and embedding into the Eclipse IDE.

```
// 1) Language specification and workspace management
data LangSpec = {
  func syntax() -> path; func startSymbol() -> string; func styling() -> path
}
data Workspace = {
  func extensions() -> string*; func langSpec(path) -> LangSpec
}
func createWorkspace(string, path) -> Workspace
func getWorkspace(root: path) -> Workspace = {
  val text = read(root + "/workspace.cfg"); createWorkspace(text, root)
}
// 2) Creating parse tables and parsing
data ParseTable {} data Ast {} data Token {} data Msg {}
func sdf2table(path) -> ParseTable
func jsglrParse(string, string, ParseTable) -> (Ast?, Token*?, Msg*)
func parse(text: string, langSpec: LangSpec) -> (Ast?, Token*?, Msg*) = {
  val mainFile = langSpec.syntax(); requires mainFile;
  val startSymbol = langSpec.startSymbol();
  val table = sdf2table(mainFile); jsglrParse(text, startSymbol, table)
}
// 3) Syntax-based styling
data SyntaxStyler {} data Styling {}
func esv2styler(path) -> SyntaxStyler
func esvStyle(Token*, SyntaxStyler) -> Styling
func style(tokens: Token*, langSpec: LangSpec) -> Styling = {
  val mainFile = langSpec.styling(); requires mainFile;
  val styler = esv2styler(mainFile); esvStyle(tokens, styler)
}
// 4) Combine parsing and styling to process strings and files
func processString(text: string, langSpec: LangSpec) -> (Msg*, Styling?) = {
  val (ast, tokens, msgs) = parse(text, langSpec);
  val styling = if(tokens != null) style(tokens, langSpec) else null;
  (msgs, styling)
}
func processFile(file: path, langSpec: LangSpec) -> (Msg*, Styling?) =
  processString(read file, langSpec)
// 5) Keep files of an Eclipse project up-to-date
func updateProject(root: path, project: path) -> (path, Msg*, Styling?)* = {
  val workspace = getWorkspace(root);
  val relevantFiles = walk project with extensions workspace.extensions();
  [updateFile(file, workspace) | file <- relevantFiles]
}
func updateFile(file: path, workspace: Workspace) -> (path, Msg*, Styling?) = {
  val langSpec = workspace.langSpec(file);
  val (msgs, styling) = processFile(file, langSpec); (file, msgs, styling)
}
// 6) Keep an Eclipse editor up-to-date
func updateEditor(text: string, file: path, root: path) -> (Msg*, Styling?) = {
  val workspace = getWorkspace(root); val langSpec = workspace.langSpec(file);
  processString(text, langSpec)
}
```

**Processing Files in the IDE**    The fourth part combines parsing and styling to process a single string or file and return the error messages and styling, which we can display in the Eclipse IDE. The fifth and sixth parts interface with the Eclipse IDE, by providing functions to keep an Eclipse project and editor up-to-date. A project is kept up-to-date by `walk`ing over the relevant files of the project, and returning the messages and styling for each file which are displayed in Eclipse. An editor is kept up-to-date by processing the text in the editor.

### 6.2  Analysis

In this section we discuss the observations we made while re-implementing the incremental software development pipeline of Spoofax in PIE. Overall, the re-implementation improves on the areas mentioned below.

**Canonical Pipeline Formalism**    The main benefit over the old pipeline of Spoofax is that the PIE re-implementation is written in a single and concise formalism that is easier to understand and maintain. The old pipeline of Spoofax is comprised of code and configuration in four different formalisms: 1) Maven Project Object Model (POM) file that describes the compilation of Java source code, 2) an incremental build system using the Pluto [12] Java API and runtime that builds language specifications, 3) a custom (partially incremental) build system for building and bootstrapping meta-languages, and 4) a custom language registry that manages multiple language specifications. Incrementality and persistence are only partially supported, and implemented and maintained explicitly.

In contrast, the PIE pipeline is specified as a single formalism in a readable, concise, and precise way, without having to implement incrementality and persistence explicitly.

**Exact (Dynamic) Dependencies**    Spoofax's old pipeline emits dependencies that are either overapproximated or underapproximated, resulting in poor incrementality and therefore longer execution times. For example, in Spoofax, changing the styling specification will trigger parsing, analysis, compilation, and styling for all editors, even though only recomputation of the styling is required (i.e., sound overapproximation). On the other hand, changing the syntax specification will not trigger reparsing of files that are not open in editors (i.e., unsound underapproximation). In the PIE pipeline, these problems do not occur because of the implicit incrementality of function calls, and the right path dependencies.

For example, the `parse` function creates several dependencies which enable incremental recomputation. When the input `text`, `mainFile` path, contents of the `mainFile`, or the `startSymbol` changes, the function is recomputed. Furthermore, the function creates a parse table, which is a long-running operation. However, because of incremental recomputation and persistence, the parse table is computed once, and after that only when the syntax specification changes.

■ **Listing 6**  Incremental performance benchmarking pipeline in PIE.

```
func main(jmhArgs: string*) -> path* = {
  val jar = build(); val pkg = "io.usethesource.criterion";
  val javaSrcDir = ./src/main/java/io/usethesource/criterion;
  val benchs: (string, string, path*)* = [ // Benchmarks name, pattern, classes
    ("set","$pkg.JmhSetBenchmarks.(*)\$",[javaSrcDir+"/JmhSetBenchmarks.java"])
  , ("map","$pkg.JmhMapBenchmarks.(*)\$",[javaSrcDir+"/JmhMapBenchmarks.java"])
  ];
  val subjs: (string, string, path*)* = [ // Subjects name, identifier, libs
    ("clojure"     , "VF_CLOJURE"     , [./lib/clojure.jar     ])
  , ("champ"       , "VF_CHAMP"       , [./lib/champ.jar       ])
  , ("scala"       , "VF_SCALA"       , [./lib/scala.jar       ])
  , ("javaslang"   , "VF_JAVASLANG"   , [./lib/javaslang.jar   ])
  , ("unclejim"    , "VF_UNCLEJIM"    , [./lib/unclejim.jar    ])
  , ("dexx"        , "VF_DEXX"        , [./lib/dexx.jar        ])
  , ("pcollections", "VF_PCOLLECTIONS", [./lib/pcollections.jar])
  ];
  [run_benchmark(jar, jmhArgs, bench, subj) | bench <- benchs, subj <- subjs]
}
func build() -> path = {
  val pomFile = ./pom.xml; requires pomFile;
  [requires file | file <- walk ./src with extensions ["java", "scala"]];
  exec(["mvn", "verify", "-f", "$pomFile"]);
  val jar = ./target/benchmarks.jar;
  generates jar; jar
}
func run_benchmark(jar: path, jmhArgs: string*, bench: (string, string, path*),
  subj: (string, string, path*)) -> path = {
  val (bName, bId, bDeps) = bench; [requires dep | dep <- bDeps];
  val (sName, sId, sDeps) = subj;  [requires dep | dep <- sDeps];
  val csv = ./results/${bName}_${sName}.csv;
  requires jar by hash;
  exec(["java", "-jar", "$jar"] + bId + ["-p", "subject=$sId"] + jmhArgs +
    ["-rff", "$csv"]);
  generates csv; csv
}
```

**Support for Complex Pipeline Patterns**   Due to space constraints, the code listing in Listing 5 omits the parts necessary for using Spoofax's name binding language and constraint solver, interfacing with existing Spoofax languages, and bootstrapping languages, but our re-implementation does support the aforementioned features. The full implementation can be found online [27].

# 7   Case Study: Live Performance Testing

In this section we evaluate PIE on a case study for continuously monitoring the performance of a set of libraries. Specifically we use a snapshot of the *Criterion* benchmark suite [39] that measures the performance of immutable hash-set/map data structures on the JVM. The snapshot of Criterion was submitted as a well-documented artifact to accompany the findings of a research paper [40].

Under the hood, Criterion uses the Java Microbenchmarking Harness (JMH) [21] to execute benchmark suites against seven data structure libraries, producing Comma-Separated Values (CSV) files with statistical-relevant benchmarking data. Criterion uses bash scripts for orchestration, requiring to re-run all benchmarks whenever a benchmark or subject library changes. Those scripts are not able to exploit incrementality, which is tedious since benchmarking all combinations takes roughly two days, to produce statistically significant outputs.

We re-engineered the pipeline such that initially each subject and benchmark combination is tested in isolation, and then incrementally re-execute all benchmarks for a particular subject if and only if that subject changes. In case the implementation of a benchmark changes, all subjects are re-tested for that benchmark. Regardless of the scenario, the CSV result files are kept up-to-date for subsequent data visualization.

We can apply such a pipeline on a local machine while developing the benchmarks for timely performance test results, or on a remote benchmarking server to minimize the amount of benchmarking work when something changes. While it is technically possible to write such an incremental pipeline in bash scripts, it would require a lot of manual work to implement, and will likely result in error-prone code. Fortunately, it is straightforward to write this pipeline in PIE.

### 7.1 Pipeline Re-Implementation

Listing 6 illustrates the benchmarking pipeline in PIE. The `build` function builds the benchmark and yields an executable JAR file `./target/benchmarks.jar`, by invoking Maven on the POM file `./pom.xml`. The `build` function `requires` all Java and Scala source files, to ensure that the JAR file is rebuilt as soon as a single source files changes.

To produce a CSV result file, the `run_benchmark` function executes the JAR file with the necessary command-line arguments for the JMH library, including the combination of `benchmark` and `subject`. The tuples `benchmark` and `subject` both store unique name identifiers —that are later used for naming the CSV file— and references to files they are comprised of. These file references are used by PIE to create dependencies for incremental re-execution.

Finally, `main` glues everything together by creating a list of benchmarks and subjects, running the benchmark with each combination of those, and by returning the up-to-date CSV files for subsequent data visualization.

### 7.2 Analysis

Compared to the existing bash script, the PIE pipeline provides incremental and persistent execution, and static analysis. The main benefit of the PIE pipeline over the bash script is that it provides incremental execution by function calls and path dependency annotations. In bash, implementing an incremental pipeline requires the pipeline developer to explicitly encode dependency tracking, change detection, caching,

■ **Table 1** Feature overview of PIE and related work (● = full support, ◐ = partial/limited support, ○ = no support).

| | Make | Automake | OMake | Tup | PROM | Nix | Maven | Ant | Gradle | Jenkins | Shake | Pluto | Fabricate | Spark | Reactive Programming | Workflow Languages | PIE |
|---|---|---|---|---|---|---|---|---|---|---|---|---|---|---|---|---|---|
| Low Boilerplate | ◐ | ◐ | ◐ | ◐ | ◐ | ◐ | ○ | ○ | ○ | ○ | ○ | ○ | ○ | ◐ | ○ | ● | ● |
| Static Analysis | ○ | ○ | ○ | ○ | ○ | ○ | ○ | ● | ○ | ○ | ● | ● | ○ | ● | ● | ◐ | ● |
| Dynamic File Dependencies | ◐ | ● | ● | ◐ | ● | ○ | ○ | ◐ | ○ | ○ | ◐ | ● | ● | ● | ○ | ○ | ● |
| Implicit Incrementality | ● | ● | ● | ● | ● | ● | ○ | ○ | ○ | ○ | ● | ○ | ● | ● | ◐ | ○ | ● |
| Embeddable | ○ | ○ | ○ | ○ | ○ | ○ | ○ | ○ | ○ | ○ | ● | ● | ● | ● | ● | ○ | ● |
| Restartable | ● | ● | ● | ● | ● | ● | ● | ● | ● | ● | ● | ● | ● | ● | ○ | ○ | ● |
| Cross-platform | ◐ | ◐ | ● | ● | ● | ○ | ● | ● | ● | ● | ● | ● | ● | ○ | ● | ● | ● |

and more, which is why the existing bash script is not incremental. In the PIE pipeline, incrementality comes from stating the `requires` and `generates` dependencies in each function, which is straightforward because it is clear what the dependencies of each function are.

Furthermore, PIE performs static name and type analysis, before executing the pipeline, whereas bash has no static checks at all. This means that errors such as simple typographical errors, or appending a value of a wrong type to a list of strings, result in a static error in PIE which is easily fixed, but result in run-time errors in bash.

## 8 Related Work

In this section we discuss related work with a focus on build systems. Table 1 provides a feature overview of the systems we discuss throughout this section.

**Partial Domain-Specific Build Abstractions**    Make [38] is a build automation tool based on declarative rules. Make extracts a static dependency graph from these rules, and executes the commands according to the dependency graph. Upon re-execution, Make is able to detect unchanged files that do not require regeneration based on time-stamps. Make supports a limited form of dynamic dependencies that does not generalize, i.e., an include directive that allows loading other Makefiles.

Automake [30] alleviates many of Make's shortcoming by introducing a formalism on top of Make that generates Makefiles. Automake is mostly geared towards C compilation and other compilation processes that follow similar patterns, but cannot be used to write arbitrary interactive pipelines, making it less flexible than PIE. Due to a lack of static checking, ill-typed Automake scripts may propagate defects — that are only detectable at run time — to the generated Makefiles. In contrast, PIE catches such errors statically before pipeline execution.

OMake [18] is a build tool with a Make-like syntax, but with a richer dependency tracking mechanism. PIE is similar to OMake in that both supports a form of dynamic path dependencies (called side-effects in OMake), and incrementality based on these dependencies. However, like Make, OMake works exclusively with files and command-line processes, meaning that it is not possible to depend on the result of a function call, or to interface with foreign functions and types, making it unsuitable for interactive pipelines which require embedding into an interactive system.

Tup [37] is a build tool with Make-like rules. Tup automatically infers required file dependencies by instrumenting the build process, providing more fine-grained dependencies than Make. However, the dependency on the input file and generated file must still be declared statically upfront.

PROM [25] is a Prolog-based make tool where Make-like builds are specified declaratively and executed as Prolog terms, increasing expressiveness and ease of development. PROM's update algorithm executes in two phases, where first a file-dependency graph is created, after which creation rules are executed to create new files, or to update out-of-date files. Because of these phases, PROM does not support dynamic discovery of dependencies during build execution.

Nix [7, 9] is a purely functional language for building and deploying software, aimed to manage configurations of the operating system NixOS [8]. Nix supports incremental execution of pipelines through cryptographic hashes of attributes and files, but must be explicitly initiated by the developer through the use of the `mkDeriva-tion` function. While incrementality becomes explicit, Nix puts the burden on pipeline developers, whereas PIE supports incrementality implicitly. Furthermore, Nix is dynamically checked, meaning that name and type defects are reported at runtime, as opposed to before runtime with static checking in PIE.

Maven [17] is a software dependency management and build tool, popular in the Java ecosystem. It features a fixed sequence of pipeline steps such as compile, package, and deploy, which are configured through an XML file. Maven is neither incremental nor interactive, requiring a full batch re-execution every time data in the pipeline changes.

Ant [16] is a build automation tool, using XML configuration files for defining software development pipelines. Ant supports incrementality by inserting `uptodate` statements that check if a source file is up to date with its target file, making incrementality explicit, at the cost of burdening the developer. Ant does not provide static analysis.

Gradle [20] is a build automation tool, programmable with the Groovy language, featuring domain-specific library functions to specify builds declaratively. Gradle supports incremental task execution through annotations that specify a task's input/output variables, files, and directories. Like Make, dependencies have to be specified statically up-front, causing an overapproximation of dependencies.

Jenkins is a continuous integration server which can be programmed with its Groovy pipeline and a set of domain-specific library functions [22]. Jenkins can detect changes to a (remote) source code repository to trigger re-execution of an entire build pipeline, however without support for incrementality.

**Software Development Pipelines as a Library**   Subsequently discussed software pipeline solutions are available as a library (i.e., internal DSL) implemented in a general-purpose programming language. Unlike an external DSL solution such as PIE, those libraries do not support domain-specific syntax or error reporting in terms of the domain, instead requiring encoding of domain concepts. Furthermore, it is hard, if not impossible, to restrict features of a programming language via a library, that heavily influence incrementality, such as mutable state. Finally, the compiler of the PIE DSL can be retargeted to a different environment or programming language, enabling a PIE pipeline to be embedded into different interactive environments without (or minimal) alteration.

Shake [33, 34] is a Haskell library for specifying build systems. Unlike Make, required file dependencies can be specified during builds in Shake, supporting more complex dependencies and reducing overapproximation of dependencies. However, like Make, targets (generated file dependencies) have to be specified up-front. This means that it is not possible to specify builds where the names of generated files are decided dynamically. For example, the Java compiler generates a class file for each inner class in a source file, where its file names are based on the inner and outer class name. Therefore, the generated file dependencies of the Java compiler are decided dynamically, and cannot be specified in Shake.

Pluto [12] is a Java library for developing incremental builds, which we have already discussed extensively in Section 2. One difference between Pluto and PIE is that Pluto supports incremental cyclic builders, whereas PIE does not. We have opted not to implicitly support cycles for simplicity of the build algorithm, and because cycles typically do not appear in pipelines. Cycles can be handled explicitly in PIE by programming the cyclic computation inside a single pipeline function.

Fabricate [19] is a Python library for developing incremental and parallel builds, that aims to automatically infer all file dependencies by tracing system calls. System call tracing is not cross-platform, only fully supporting Linux at the moment. PIE in contrast is cross-platform, because its runtime works on any operating system the JVM runs on.

Apache Spark [4] is a big data processing framework where distributed datasets are transformed by higher-order functions. PIE is similar to Spark in that both create dependency graphs between calculations. For example, when transforming a dataset with Spark (e.g., with map or filter), the derived dataset depends on the parent dataset, such that the derived dataset is rederived when the parent changes. PIE differs from Spark in that PIE works with local data only, whereas Spark works with a distributed storage system required for big data processing. However, PIE supports arbitrary computations (as opposed to a fixed set of higher-order functions in Spark), and dynamic file dependencies.

**General-Purpose Languages**   Reusing an existing general-purpose language, such as Java or Haskell, and giving it an incremental and persistent interpretation is not feasible for several reasons. It requires adding additional constructs to the language, such as path dependencies and operations, which require changes to the syntax, static semantics, dynamic semantics (compiler or interpreter) of the language. That requires at the very least being able to change the language, which is not always possible. Even when it is possible, language parsers, checkers, and compilers are often large codebases that require significant effort to change. Furthermore, we also need to ensure that existing constructs work under incrementality. For example, mutable state in Java interferes with incrementality.

**Reactive Programming**    Reactive programming is characterized by asynchronous data stream processing, where data streams form a pipeline by composing streams with a set of stream combinators. Reactive programming approaches come in the form of libraries implemented in general-purpose languages, such as Reactive Extensions [31], or as an extended language such as REScala [36]. Reactive programming approaches provide a form of incrementality where the reactive pipeline will rerun if any input signal changes. However, they do not cache outputs or prevent re-execution of pipeline steps when there are no changes. Note that reactive programming approaches operate in volatile memory only, whereas PIE's runtime supports persistence (i.e., pause and resume) of pipeline executions. Preserving a pipeline's state is of special importance in interactive environments such as IDEs, to support restarting the programming environment without re-triggering potentially expensive calculations. Furthermore, reactive programming approaches do not support (dynamic) file dependencies.

**Workflow Languages**    Workflows, like pipelines, describe components (processors) and how data flows between these components. Workflow languages are DSLs which are used to model data analysis workflows [2], business process management [1], and model-to-model transformations [5], among others. The crucial differentiation between many workflow systems and software development pipelines, is that the former model manual steps that require human interaction, whereas the latter focuses on processors that perform general purpose computations.

## 9    Future Work

We now discuss directions for future work.

**First-Class Functions and Closures**    Currently, the PIE DSL does not support first-class functions and closures, for simplicity. The PIE runtime does support first-class functions, since function calls are immutable and serializable values which can be passed between functions and called. However, closures are not yet supported, because (again for simplicity), functions must be registered with the runtime before a pipeline is executed.

To fully support first-class functions and closures, we must add them to the PIE DSL, and support closures in the runtime and API. This requires closures to be serializable, which the JVM supports. Closures from foreign functions must ensure not to capture mutable state, non-serializable values, or large objects graphs, as these can break incrementality. Spores [32] could be used to guarantee these properties for closures.

**Live Pipelines**   PIE pipelines can dynamically evolve through the inputs into the pipeline: files on the filesystem such as configuration files, and objects passed through function calls such as editor text. However, the pipeline code itself currently cannot dynamically evolve at runtime. When the pipeline code itself is changed, the pipeline must be recompiled and reloaded. This process is relatively fast, because compiling PIE DSL code and restarting the JVM is fast, but can be improved nevertheless. Furthermore, dynamic evolution of pipelines at runtime is especially important if we want to apply PIE to live programming environments.

While there are known solutions for compiling and reloading code in the JVM, such as using class loaders, it is unclear how to handle incrementality in the face of changes to the pipeline program. For example, if the `normalize` function in Listing 2 is changed, all calls of `normalize` are potentially out-of-date and need to be re-executed, as well as all function calls that (transitively) call `normalize`. Similarly, foreign functions and data types can be changed, which require re-execution or even data migrations in the persistent storage.

## 10 Conclusion

We have presented PIE: a DSL, API, and runtime for developing interactive software development pipelines. PIE provides a straightforward programming model that enables direct and concise expression of pipelines with minimal boilerplate, reducing the development and maintenance effort of pipelines. Compared to the state of the art, PIE reduces the code required to express an interactive pipeline by a factor of 6 in a case study on syntax-aware editors. Furthermore, we have evaluated PIE on two complex interactive software development pipelines, showing that the domain-specific integration of features in PIE enable concise expression of pipelines, which are normally cumbersome to express with a combination of traditional build systems and general-purpose languages.

### Acknowledgements

This research was supported by NWO/EW Free Competition Project 612.001.114 (Deep Integration of Domain-Specific Languages) and NWO VICI Project (639.023.206) (Language Designer's Workbench).

## About the authors

**Gabriël Konat** is a PhD candidate at Delft University of Technology, where he works on domain-specific languages, language workbenches, bootstrapping, and incrementalization. His current work includes designing a DSL for programming incremental pipelines, improving the scalability of incremental pipeline algorithms, and applying incremental pipelines to language workbenches. You can contact him at g.d.p.konat@tudelft.nl.

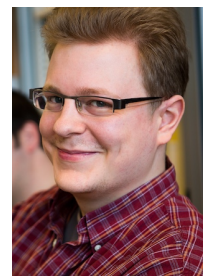

**Michael J. Steindorfer** is a senior software engineer working in industry and a guest researcher at the Delft University of Technology. His research and engineering efforts focus on optimizing functional data structures, the design and implementation of programming languages, and improving big data processing runtimes for cloud infrastructures. You can contact him at michael@steindorfer.name.

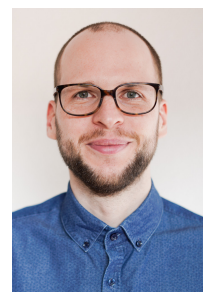

**Sebastian Erdweg** is an assistant professor at Delft University of Technology, where he works on the foundation and application of programming languages. His current work includes incremental static analysis and build systems, modernizing legacy code to adopt new language features, and safe refactorings, analyses, type systems, and program transformations. You can contact him at s.t.erdweg@tudelft.nl and find further information at http://erdweg.org.

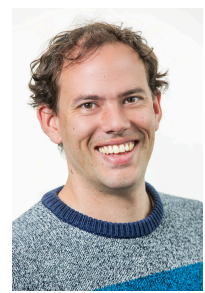

**Eelco Visser** is Antoni van Leeuwenhoek Professor of Computer Science and chair of the Programming Languages Group at Delft University of Technology. His current research is on the foundation and implementation of declarative specification of programming languages. You can contact him at e.visser@tudelft.nl and find further information at http://eelcovisser.org.

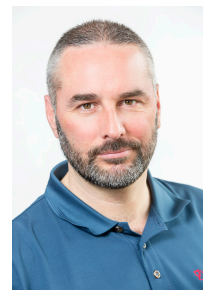